\documentclass[12pt]{amsart}
\usepackage{amsmath,amssymb}

\newcommand{\RR}{\mathbb{R}}
\newcommand{\LL}{\mathcal{L}}

\begin{document}

\title{New variables for the Einstein theory of gravitation}
\author{L.~D.~Faddeev}
\address{St.Petersburg Department of Steklov Mathematical Institute}

\begin{abstract}
    The set of 10 covariant vector fields
$ f_{\mu}^{A}(x) $
    is taken as basic variables to describe the gravitational field.
    Metric 
$ g_{\mu\nu} $
    is a composite field. A possibility for the gravitational constant
    to be described as a condensate of additional scalar field is discussed.
\end{abstract}

\maketitle

\section{Introduction}
    Newton gravitational constant
$ \kappa $ in units
$ \hbar=1 $, $ c=1 $,
    entering the Hilbert-Einstein lagrangian has dimension of mass
\begin{equation*}
    [\frac{1}{\kappa^{2}}] = [\text{M}]^{2} .
\end{equation*}
    This leads to the nonrenormalisability of quantum theory.
    The similar problem of massive vector field, describing weak interaction,
    was solved in Weinberg-Salam theory
\cite{Weinberg},
\cite{Salam}
    by adding the Higgs field. It is temptating to look for the simlar
    solution in the Einstein theory of gravitation.
    There were numerous attempts to realize such a programm, beginning with
    Brans and Dicke
\cite{Brans},
    see also a detailed discussion in
\cite{Adler}.
    Here I present one more proposal in this direction.

    To make the similarity more close
    one should try to describe gravity by means of vector fields.

    Vector fields were used in gravitation theory in the framework of the
    embedding approach
\cite{Rumer}--\cite{PF}.
    The 4 dimensional space-time
$ M_{4} $
    is described as a hypersurface in 10-dimensional linear space
$ \RR^{10} $.
    The coordinates
$ f^{A} $, $ A=1,\ldots 10$
    become functions
$ f^{A}(x) $
    of coordinates
$ x^{\mu} $, $ \mu=1,\ldots 4$
    on 
$ M_{4} $
    and metric
$ g_{\mu\nu} $
    is an induced one
\begin{equation*}
    g_{\mu\nu}(x) = \partial_{\mu}f^{A}(x) \partial_{\nu}f^{A}(x).
\end{equation*}
    Here and in the following I use only euclidean signature for simplicity.
    Thus the covariant vector fields
\begin{equation}
\label{CVF}
    f_{\mu}^{A}(x) = \partial_{\mu} f^{A}(x)
\end{equation}
    appear as the basic variables.
    The curvature tensor, given by Gauss formula,
\begin{equation*}
    R_{\alpha\beta,\mu\nu} = \Pi^{AB} (
	\partial_{\mu} f_{\alpha}^{A} \partial_{\nu }f_{\beta}^{B}
	- \partial_{\nu}f_{\alpha}^{A} \partial_{\mu} f_{\beta}^{B}) ,
\end{equation*}
    where
$ \Pi^{AB} $
    is a projector to vertical direction
\begin{equation*}
    \Pi^{AB} = \delta^{AB} - g^{\lambda\sigma} f_{\lambda}^{A} f_{\sigma}^{B},
\end{equation*}
    is quadratic in these vector fields.
    However the appearance of the derivative in the definition of
$ f_{\mu}^{A}(x) $
    leads to equations of motion, different from Hilbert-Einstein equations.
    Indeed,
\begin{equation*}
    \delta \int \sqrt{g} R d^{4}x =
	2 \int \sqrt{g} \partial_{\mu}(G^{\mu\nu}f_{\nu}^{A}) \delta f^{A}
	    d^{4}x ,
\end{equation*}
    where
$ G^{\mu\nu} = R^{\mu\nu} - \frac{1}{2} g^{\mu\nu} R $,
    and we see, that 
$ G^{\mu\nu} $
    enters the equations of motion with extra derivative.
    This fact is considered as drawback in
\cite{RT},
    some remedy is discussed in
\cite{PF}.

    In my recent note
\cite{LDF}
    I proposed more radical solution, namely I took the fields
$ f_{\mu}^{A}(x) $
    as a genuine covariant vector fields and forgot the embedding.
    It was shown in
\cite{LDF},
    that in this way one gets proper Hilbert-Einstein equations.
    One of features of approach in
\cite{LDF}
    was introduction of constraints
\begin{equation}
\label{cons}
    \partial_{\mu}f_{\nu}^{A} - \partial_{\nu} f_{\mu}^{A} = 0
\end{equation}
    which led to
(\ref{CVF})
    by means of adding the tensor lagrange multiplier
$ B^{\mu\nu,A} $.
    However the more detailed analysis of equations of motion in
\cite{LDF},
    which was done by S.~Paston
\cite{Paston},
    showed that this trick is not needed.
    The improved variant of the approach of
\cite{LDF}
    is given here in the section 3.
    It shows complete equivalence with Hilbert-Einstein equations of motion
    without using
    extra condition
(\ref{cons}).

    In section 4 I discuss some possibilities to explore the condensate idea.
    Additional scalar fields
$ X^{A} $
    are introduced and several proposals for the corresponding lagrangian are
    discussed.

    The work on this project was supported in part by the
RFBR grant 08-01-00638 and programm ``Mathematical
problems of nonlinear dynamics'' of Russian Academy of Sciences.

I am greatful to S.~Paston for important comments.

\section{Differential geometry}
    Here I shall introduce some formulas from differential geometry, which
    will be used in section 3.

    Let 
$ M_{4} $
    be 4-dimensional space-time with coordinates
$ x^{\mu} $, $ \mu = 1,\ldots 4$.
    Consider a set of covariant vector fields
$ f_{\mu}^{A} $, $ A = 1,\ldots D$.
    The number
$ D $
    is large enough, later we shall see that the natural choice is
$ D=10 $.
    The two index field
\begin{equation*}
    g_{\mu\nu} = f_{\mu}^{A} f_{\nu}^{A}
\end{equation*}
    and three index field
\begin{equation*}
    \Omega_{\alpha,\beta\mu} = f_{\alpha}^{A} \partial_{\mu} f_{\beta}^{A}
\end{equation*}
    define  metric and linear connection on 
$ M_{4} $.
    Indeed under the coordinate transformation
\begin{equation*}
    \delta x^{\mu} = \epsilon^{\mu}(x)
\end{equation*}
    we have by definition of covariant vector field
\begin{equation*}
    \delta f_{\mu}^{A} = -\partial_{\mu}\epsilon^{\lambda} f_{\lambda}^{A}
	- \epsilon^{\lambda} \partial_{\lambda} f_{\mu}^{A}, 
\end{equation*}
    so that
$ g_{\mu\nu} $
    transforms as a tensor and in the transformation of
$ \Omega_{\alpha,\beta\mu} $
    besides the usual tensorial terms we have noncovariant contribution
$ g_{\alpha\lambda}\partial_{\beta}\partial_{\mu}\epsilon^{\lambda} $
    characteristic of linear connection.

    Covariant indexes are raised as usual by the inverse matrix
$ g^{\mu\nu} $
\begin{equation*}
    g^{\mu\rho}g_{\rho\nu} = \delta^{\mu}_{\nu} , \quad
	f^{\mu A} = g^{\mu\nu} f_{\nu}^{A} .
\end{equation*}
    The covariant derivative of covariant vector field
$ X_{\mu} $
    is given by
\begin{equation*}
    \nabla_{\mu} X_{\nu} = \partial_{\mu}X_{\nu}
	-\Omega_{\nu\mu}^{\rho} X_{\rho} ,
\end{equation*}
    where
\begin{equation*}
    \Omega_{\nu\mu}^{\rho} = g^{\rho\alpha} \Omega_{\alpha,\nu\mu}
	= f^{\rho A} \partial_{\mu}f_{\nu}^{A} .
\end{equation*}
    
    Connection
$ \Omega_{\beta\mu}^{\alpha} $
    is compatible with metric
$ g_{\mu\nu} $.
    Indeed
\begin{multline*}
    \nabla_{\lambda} g_{\mu\nu} = \partial_{\lambda} f_{\mu}^{A} f_{\nu}^{A}
	+ f_{\mu}^{A} \partial_{\lambda} f_{\nu}^{A} -
	\Omega_{\mu\lambda}^{\sigma} g_{\sigma\nu} -
	\Omega_{\nu\lambda}^{\sigma} g_{\mu\sigma} = \\
    = \partial_{\lambda}f_{\mu}^{A}f_{\nu}^{A} 
	+ f_{\mu}^{A} \partial_{\lambda} f_{\nu}^{A} -
    \partial_{\lambda}f_{\mu}^{A} f^{\sigma A}g_{\sigma\nu} -
    \partial_{\lambda}f_{\nu}^{A} f^{\sigma A}g_{\mu\sigma} = 0,
\end{multline*}
    however it has torsion
\begin{equation*}
    T_{\alpha,[\mu\nu]} = \Omega_{\alpha,\mu\nu} - \Omega_{\alpha,\nu\mu} .
\end{equation*}
    The riemanian connection
$ \Gamma_{\alpha,\beta\mu} $ 
    is expressed via
$ \Omega_{\alpha,\beta\mu} $
    as follows
\begin{equation*}
    \Gamma_{\alpha,\beta\mu} =
	\frac{1}{2}(\Omega_{\alpha,\beta\mu}+\Omega_{\alpha,\mu\beta}) 
	+ \frac{1}{2}(\Omega_{\mu,\alpha\beta}-\Omega_{\mu,\beta\alpha})
	+ \frac{1}{2}(\Omega_{\beta,\alpha\mu}-\Omega_{\beta,\mu\alpha})
\end{equation*}
    and so
$ \Gamma_{\alpha,\beta\mu} =  \Omega_{\alpha,\beta\mu} $
    under the conditions of vanishing torsion
$    T_{\alpha,[\mu\nu]} $.

    Let us calculate the curvature tensor
$ S_{\mu\nu,\beta}^{\alpha} $ of connection
$ \Omega $.
    We have by definition
\begin{multline*}
    S_{\beta,\mu\nu}^{\alpha} = \partial_{\mu} \Omega_{\beta\nu}^{\alpha}
	- \partial_{\nu} \Omega_{\beta\mu}^{\alpha} 
	+ \Omega_{\sigma\mu}^{\alpha} \Omega_{\beta\nu}^{\sigma}
	- \Omega_{\sigma\nu}^{\alpha} \Omega_{\beta\mu}^{\sigma} = 
    \partial_{\mu} (f^{\alpha A}\partial_{\nu} f_{\beta}^{A}) - \\
    - \partial_{\nu} (f^{\alpha A}\partial_{\mu} f_{\beta}^{A})
    + f^{\alpha A}\partial_{\mu}f_{\sigma}^{A} f^{\sigma B}
	\partial_{\nu}f_{\beta}^{B}
    - f^{\alpha A}\partial_{\nu}f_{\sigma}^{A} f^{\sigma B}
	\partial_{\mu}f_{\beta}^{B} .
\end{multline*}

    The second derivatives cancel and using elementary property
\begin{equation*}
    f^{\alpha A} \partial_{\mu}f_{\sigma}^{A} =
	- \partial_{\mu} f^{\alpha A} f_{\sigma}^{A} ,
\end{equation*}
    which follow from orthonormality
$ f_{\mu}^{A}f^{A\nu} = \delta_{\mu}^{\nu} $,
    we can rewrite 
$ S_{\mu\nu,\beta}^{\alpha} $ 
    as
\begin{equation*}
    S_{\beta,\mu\nu}^{\alpha} = \Pi^{AB}
	(\partial_{\mu}f^{\alpha A}\partial_{\nu}f_{\beta}^{B} 
	- \partial_{\nu}f^{\alpha A}\partial_{\mu}f_{\beta}^{B} ) ,
\end{equation*}
    where
\begin{equation*}
    \Pi^{AB} = \delta^{AB} - f_{\sigma}^{A} f^{\sigma B}
\end{equation*}
    is a projector, orthogonal to vector fields
$ f_{\mu}^{A} $.
    Now we can lower the index
$ \alpha $
    to get
\begin{equation*}
    S_{\alpha\beta,\mu\nu} = \Pi^{AB} (
	\partial_{\mu}f_{\alpha}^{A} \partial_{\nu}f_{\beta}^{B}
	- \partial_{\nu}f_{\alpha}^{A} \partial_{\mu}f_{\beta}^{B} ).
\end{equation*}
    Indeed
\begin{equation*}
    g_{\alpha \rho} \partial_{\mu} f^{\rho A} = 
	\partial_{\mu} ( g_{\alpha \rho} f^{\rho A}) -
	\partial_{\mu} g_{\alpha\rho} f^{\rho A}
\end{equation*}
    and second term is killed by the projector.
    Now using the property of projector
\begin{equation*}
    \Pi^{AB} = \Pi^{AC} \Pi^{BC}
\end{equation*}
    we can finally rewrite
$ S_{\mu\nu,\alpha\beta} $
    as
\begin{equation*}
    S_{\alpha\beta,\mu\nu} = b_{\alpha\mu}^{A} b_{\beta\nu}^{A}
	- b_{\alpha\nu}^{A} b_{\beta\mu}^{A}
\end{equation*}
    via covariant tensor fields
\begin{equation*}
    b_{\alpha\mu}^{A} = \Pi^{AB} \partial_{\mu} f_{\alpha}^{B} .
\end{equation*}
    We see, that
$ \Omega_{\alpha\mu}^{\beta} $ and
$ b_{\alpha\mu}^{A} $
    constitute ```horisontal'' and ``vertical'' components of
$ \partial_{\mu}f_{\alpha}^{A} $.
    The last formula for the
$ S_{\alpha\beta,\mu\nu} $
    looks similarly to the Gauss formula, expressing the curvature tensor of
    embedded manifold via set of second quadratic forms.
    However our tensors
$ b_{\alpha\mu}^{A} $
    are not symmetric
\begin{equation*}
    b_{\alpha\mu}^{A} \ne b_{\mu\alpha}^{A} .
\end{equation*}
    Thus the curvature tensor
$ S_{\alpha\beta,\mu\nu} $
    is antisymmetric with respect to interchange
$ \mu \leftrightarrow \nu $,
$ \alpha \leftrightarrow \beta $,
    but not symmetric to
$ \mu\nu \leftrightarrow \alpha \beta $.

\section{Variational equations}
    The lagrangian
\begin{align*}
    \LL &= \sqrt{g} g^{\mu\alpha} g^{\nu\beta} S_{\alpha\beta,\mu\nu} = \\
    & = \sqrt{g} \Pi^{AB} (
	\partial_{\mu} f^{\mu A} \partial_{\nu}f^{\nu B}
	- \partial_{\nu} f^{\mu A} \partial_{\mu}f^{\nu B} ) .
\end{align*}
    was already considered in
\cite{LDF}.
    We shall not add to it the second term
$ \sqrt{g} B^{\mu\nu,A}(\partial_{\mu}f_{\nu}^{A} -\partial_{\nu}f_{\mu}^{A}) $.
    Indeed, as was realized after the publication of
\cite{LDF},
    the important comment by S.~Paston
\cite{Paston}
    shows, that this is not needed.

    The variation
\begin{equation*}
    \delta \int \LL d^{4}x = \int \Sigma_{\alpha}^{A} \delta f^{\alpha A}
	d^{4} x
\end{equation*}
    has the following form
\begin{equation*}
    \Sigma_{\alpha}^{C} = \sqrt{g} \bigl(2Q_{\mu}^{ABC}S_{\alpha}^{\mu AB}
	- Q_{\alpha}^{ABC} S^{AB} \bigr),
\end{equation*}
    where
\begin{align*}
    Q_{\mu}^{ABC} =& \Pi^{AB}f_{\mu}^{C} + \Pi^{AC}f_{\mu}^{B}
	+ \Pi^{BC}f_{\mu}^{A}, \\
    S_{\alpha}^{\mu AB} =& \partial_{\alpha}f^{\mu A} \partial_{\nu}f^{\nu B}
	- \partial_{\nu} f^{\mu A} \partial_{\alpha} f^{\nu B} 
\end{align*}
    and
\begin{equation*}
    S^{AB} = \partial_{\mu}f^{\mu A} \partial_{\nu} f^{\nu B}
    - \partial_{\nu}f^{\mu A} \partial_{\mu} f^{\nu B} .
\end{equation*}
    Consider separately the ``vertical'' and ``horisontal'' contributions to
$ \Sigma_{\alpha}^{A} $
\begin{equation*}
    \Sigma_{\alpha}^{A} = 2\sqrt{g} \bigl(f^{\mu A} H_{\alpha\mu}
	+ \Pi^{AB} V_{\alpha}^{B} \bigr).
\end{equation*}
    The vertical part
$ V_{\alpha}^{A} $
    is given by
\begin{equation*}
    V_{\alpha}^{A} = \sqrt{g} \bigl( (\Pi^{AC}f_{\mu}^{B}+\Pi^{AB}f_{\mu}^{C})
	S_{\alpha}^{\mu BC} - \Pi^{AC} f_{\alpha}^{B} S^{BC} \bigr) 
\end{equation*}
    and using the definitions of
$ b_{\alpha\mu}^{A} $ and
$ \Omega_{\beta\mu}^{\alpha} $
    can be rewritten as follows (this was done be S.~Paston
\cite{Paston})
\begin{align*}
    V_{\alpha}^{A} = & 
	b_{\beta}^{\beta A} (\Omega_{\mu\alpha}^{\mu}-\Omega_{\alpha\mu}^{\mu})
    + b_{\alpha}^{\beta A} (\Omega_{\beta\mu}^{\mu}-\Omega_{\mu\beta}^{\mu})
    + b_{\sigma}^{\beta A} (\Omega_{\alpha\beta}^{\sigma}
	- \Omega_{\beta\alpha}^{\sigma})
\end{align*}
    and so is a linear combination of torsion. We have 24 components
    in torsion and
$ 4(D-4) $
    equtions of motion
\begin{equation}
\label{EoM}
    V_{\alpha}^{A} = 0.
\end{equation}
    Here we take into account, that due to the presence of projector in the
    definition of
$ b_{\alpha\mu}^{A} $ index
$ A $
    effectively runs through
$ D-4 $
    values.


    The set of equations
(\ref{EoM})
    is linear in the components of torsion and have the form
\begin{equation*}
    K_{\alpha}^{A}{}_{\lambda}^{[\mu,\nu]} T_{[\mu,\nu]}^{\lambda} = 0,
\end{equation*}
    where matrix elements of matrix
$ K $
    are expressed via
$ b_{\alpha}^{\beta A} $.
    In case
$ D=10 $ $ K $
    will be quadratic 
$ 24 \times 24 $
    matrix.
    We shall not write its explicit form via 96 components of 
$ b_{\alpha}^{\beta A} $,
    however we can show, that in the generic situation this matrix is
    nondegenerate.
    Thus the equations
(\ref{EoM})
    lead to the vanishing of torsion
\begin{equation*}
    T_{[\alpha\beta]}^{\mu} = 0
\end{equation*}
    and so
\begin{equation*}
    \Omega_{\beta\mu}^{\alpha} = \Gamma_{\beta\mu}^{\alpha} .
\end{equation*}

    Of course the equations
$ V_{\alpha}^{A} = 0 $
    can have another solution, imposing conditions on
$ b_{\alpha}^{\beta A} $,
    so I should present arguments, supporting the solution taken here.
    Until now I have only esthetic reasons, more serious discussion must be
    based on global variational considerations.

    The horisontal part of
$ \Sigma_{\alpha}^{A} $
\begin{equation*}
    H_{\alpha\mu} = \sqrt{g}\Pi^{BC} (S_{\alpha\mu}^{BC} 
	- \frac{1}{2} g_{\alpha\mu} S^{BC})
\end{equation*}
    give 16 equations of motion,
\begin{equation*}
    S_{\alpha\mu} - \frac{1}{2} g_{\alpha\mu} S = 0 ,
\end{equation*}
    out of which 6 are trivialy satisfied and 10 coincide with
    Hilbert-Einstein equations due to condition
\begin{equation*}
    \Omega_{\beta\mu}^{\alpha} |_{T=0} = \Gamma_{\beta\mu}^{\alpha} .
\end{equation*}
    Thus it is shown, that vector variables
$ f_{\alpha}^{A} $
    with 40 components lead to the same equations as 10 components of metric
$ g_{\mu\nu} $.

    Now I should employ new ooportunity to modify the Einstein theory of
    gravity.

\section{Scalar field and $ O(6) $ symmetry}
    Let us add to the list of fields the set of scalar fields
$ X^{A} $
    and consider separately their horisontal and vertical components. For that
    it is convenient to introduce the set of orthonormal vectors
$ e_{a}^{A} $, $ a=1,\ldots 6 $,
    orthogonal to
$ f_{\mu}^{A} $,
    considered as vectors in
$ \RR^{10} $
\begin{gather*}
    f_{\mu}^{A} e_{a}^{A} = 0 , \quad \mu=1,\ldots 4, \; a=1,\ldots 6 \\
    e_{a}^{A} e_{b}^{A} = \delta_{ab} , \quad
    \Pi^{AB} = e_{a}^{A} e_{a}^{B} .
\end{gather*}
    The horisontal components of
$ X^{A} $
    give vector field
$ Y_{\mu} = X^{A}f_{\mu}^{A} $ on
$ M_{4} $
    and vertical components
$ Z_{a} = X^{A} e_{a}^{A} $
    give vector in 
$ \RR^{6} $.
    We can introduce the connection
\begin{equation*}
    \theta_{\mu,ab} = e_{a}^{A} \partial_{\mu} e_{b}^{A}
\end{equation*}
    and so we define covariant derivatives for
$ Y_{\mu} $ and 
$ Z_{a} $
\begin{align*}
    \nabla_{\mu}^{\Omega} Y_{\alpha} & = \partial_{\mu} Y_{\alpha} -
	\Omega_{\alpha\mu}^{\beta} Y_{\beta} , \\
    \nabla_{\mu}^{\theta} Z_{a} & = \partial_{\mu} Z_{a} + \theta_{\mu,ab}Z_{b}
\end{align*}
    and use them for construction of lagrangians covariant with respect to
    coordinate transformations and local
$ O(6) $
    relations.
    It is instructive to observe, that in the natural kinetic term for the
    vertical components
\begin{equation*}
    \LL^{Z} = \sqrt{g} g^{\mu\nu} \nabla_{\mu}^{\theta}Z_{a}
	\nabla_{\nu}^{\theta} Z_{a} .
\end{equation*}
    the reference to the basis
$ e_{a}^{A} $
    disappears
\begin{equation*}
    \LL^{Z} = \sqrt{g} g^{\mu\nu} \Pi^{AB} 
	(\partial_{\mu}X^{A} - b_{\sigma\mu}^{A} Y^{\sigma})
	(\partial_{\nu}X^{B} - b_{\rho\nu}^{B} Y^{\rho}) .
\end{equation*}
    This will be shown in Appendix I.
    Also it is useful to take into account, that
$ \nabla_{\mu}^{\Omega} Y_{\alpha} $
    can be expressed via
$ b_{\alpha\mu}^{A} $
    as follows
\begin{equation}
\label{nY}
    \nabla_{\mu}^{\Omega} Y_{\nu} = f_{\nu}^{A} \partial_{\mu} X^{A}
	+ b_{\nu\mu}^{A} X^{A} ,
\end{equation}
    so that in covariant derivatives the vertical and horisontal components of
$ X^{A} $
    interchange.

    It is time to consider the dimensional attribute to all these fields.
    It is natural to consider the basic fields
$ f_{\mu}^{A} $
    dimensionless, then the metric
$ g_{\mu\nu} $
    and connection
$ \Omega_{\alpha,\beta\mu} $
    have usual dimensions
\begin{equation*}
    [g_{\mu\nu}] = [\text{L}]^{0} , \quad 
	[\Omega_{\alpha,\beta\mu}] = [\text{L}]^{-1} .
\end{equation*}
    The scalar field
$ X^{A} $
    has dimension as usual
\begin{equation*}
    [X^{A}] = [\text{L}]^{-1} .
\end{equation*}
    In this way
$ \LL^{Z} $
    and all lagrangians below have dimension
$ [\text{L}]^{-4} $,
    so that the associated coupling constants are dimensionless.

    Now we turn to the main speculation of this paper. We can introduce the
    covariant interaction lagrangian
\begin{equation}
\label{LXf}
    \LL^{X,f} = \sqrt{g} g^{\mu\alpha}g^{\nu\beta} X^{A} (
	b_{\alpha\mu}^{A} b_{\beta\nu}^{B} - b_{\alpha\nu}^{A}b_{\beta\mu}^{B})
	X^{B}
\end{equation}
    with dimension
$ [\text{L}]^{-4} $.
    Under the hypothese, that we have in quantum theory nontrivial condensate
\begin{equation*}
    <X^{A} X^{B}> = \frac{1}{\kappa^{2}} P^{AB}
\end{equation*}
    we can imply, that our version of the Hilbert-Einstein lagrangian
    reappears as an effective lagrangian in some more microscopic theory.
    Note, that this condensate does not contribute to
$ \LL^{Z} $
    because terms without derivatives there contain only
    horisontal component of
$ X^{A} $, namely
$ Y_{\mu} = f_{\mu}^{A} X^{A} $.

    It would be nice to get
$ \LL^{X,f} $
    as ``massive'' term for vector field. Indeed the natural choice
\begin{equation*}
    \hat{\LL} = \sqrt{g} g^{\mu\alpha}g^{\nu\beta} \bigl(
	\nabla_{\mu}^{\Omega}Y_{\alpha} \nabla_{\nu}^{\Omega}Y_{\beta}
	- \nabla_{\nu}^{\Omega}Y_{\alpha} \nabla_{\mu}^{\Omega}Y_{\beta}
    \bigr)
\end{equation*}
    contains the term
$ \LL^{Xf} $
    as quadratic form without derivatives due to
(\ref{nY}),
    but it will be shown in Appendix II, that it is not compatible
    with the expected condensate.

    Thus a lot of work is still ahead to explore fully all possibilities
    of vector variables to describe gravity and develope the cprresponding
    quantum theory.

\section*{Appendix I}
    We shall show here, that in the expression
$ \nabla_{\mu}^{\theta}Z_{a} \nabla_{\nu}^{\theta} Z_{a} $
    the reference to variables
$ e_{a}^{A} $
    disappear. We have
\begin{align*}
    \nabla_{\mu}^{\theta} Z_{a} & = \nabla_{\mu}^{\theta} (e_{a}^{A} X^{A})
    = e_{a}^{A} \partial_{\mu} X^{A} + (\nabla_{\mu}^{\theta}e_{a}^{A})X^{A}\\
	&= (\partial_{\mu}e_{a}^{A} + \theta_{\mu,ab}e_{b^{A}}) X^{A}
	    + e_{a}^{A} \partial_{\mu} X^{a} .
\end{align*}
    Consider the expression
$ \Pi^{AC} \partial_{\mu} \Pi^{CB} $.
    Using the definition
\begin{equation*}
    \Pi^{AB} = e_{a}^{A} e_{a}^{B}
\end{equation*}
    we get
\begin{equation*}
    \Pi^{AC} \partial_{\mu} \Pi^{CB} = e_{a}^{A}e_{a}^{C}
	(\partial_{\mu}e_{b}^{C}e_{b}^{B} + e_{b}^{C}\partial_{\mu}e_{b}^{B})
    = e_{a}^{A} \theta_{\mu,ab} e_{b}^{B} + e_{a}^{A}\partial_{\mu}e_{a}^{B}
	= e_{a}^{A} \nabla_{\mu}^{\theta} e_{a}^{B} .
\end{equation*}
    On the other hand we have
\begin{equation*}
    \Pi^{AC} \partial_{\mu} \Pi^{CB} = -\Pi^{AC}
    (\partial_{\mu}f_{\sigma}^{C}f^{\sigma B} 
	+ f_{\sigma}^{C}\partial_{\mu} f_{\sigma}^{B})
	= -f^{\sigma B} \Pi^{AC} \partial_{\mu} f_{\sigma}^{C}
	= - b_{\sigma\mu}^{A} f^{\sigma B} .
\end{equation*}
    Thus
\begin{equation*}
    e_{a}^{A} \nabla_{\mu}^{\theta} Z_{a} = \Pi^{AC} \partial_{\mu}
	\Pi^{CB} X^{B} + \Pi^{AB}\partial_{\mu} X^{B} =
    \Pi^{AB} \partial_{\mu}X^{B} - b_{\sigma\mu}^{A} Y^{\sigma} .
\end{equation*}
    Finally due to orthonormality of
$ e_{a}^{A} $
    we have
\begin{equation*}
    e_{a}^{A} \nabla_{\mu}^{\theta} Z_{a} e_{b}^{A} \nabla_{\nu}^{\theta}Z_{b}
	= \nabla_{\mu}^{\theta} Z_{a} \nabla_{\nu}^{\theta} Z_{a} ,
\end{equation*}
    from which the formula in section 4 follows.

\section*{Appendix II}
    To make analogy with WS theory one could consider the lagrangian
\begin{multline*}
    (f^{\mu A}\partial_{\mu}X^{A} + b_{\mu}^{\mu A}X^{A})
    (f^{\nu B}\partial_{\nu}X^{B} + b_{\nu}^{\nu B}X^{B}) - \\
    - (f^{\nu A}\partial_{\mu}X^{A} + b_{\mu}^{\nu A}X^{A})
    (f^{\mu B}\partial_{\nu}X^{B} + b_{\nu}^{\mu B}X^{B}) .
\end{multline*}
    Indeed, coefficients in the quadratic form
$ X^{A}X^{B} $
    coincides with the
$ \LL^{X,f} $.
    However, as already mentioned in the main text,
    the last expression is equal to
\begin{equation*}
    \sqrt{g} \bigl(
	\nabla_{\mu}Y^{\mu} \nabla_{\nu} Y^{\nu}
	    - \nabla_{\mu} Y^{\nu} \nabla_{\nu} Y^{\mu}  \bigr) 
    = \sqrt{g} \LL
\end{equation*}
    and it contains only horisontal components of 
$ X^{a} $.
    It can be transformed as follows. Denote
\begin{equation*}
    W^{\mu} = Y^{\mu} \nabla_{\nu} Y^{\nu} - Y^{\nu} \nabla_{\nu} Y^{\mu} .
\end{equation*}
    We have
\begin{equation*}
    \nabla_{\mu}W^{\mu} = \LL + Y^{\mu} \nabla_{\mu}\nabla_{\nu} Y^{\nu}
	- Y^{\nu} \nabla_{\mu}\nabla_{\nu} Y^{\mu}
	= \LL + Y^{\mu} [\nabla_{\mu},\nabla_{\nu}] Y^{\nu} .
\end{equation*}
    We can see from the expression of
$ \Gamma_{\alpha,\beta\mu} $ via
$ \Omega_{\alpha,\beta\mu} $
    that
\begin{equation*}
    \Omega_{\nu\mu}^{\mu} = \Gamma_{\mu\nu}^{\mu} + T_{\nu\mu}^{\mu}
\end{equation*}
    so that
\begin{equation*}
    \sqrt{g} \nabla_{\mu} W^{\mu} - \sqrt{g} T_{\alpha\mu}^{\alpha} W^{\mu}
	= \partial_{\mu} (\sqrt{g} W^{\mu})
\end{equation*}
    is a pure divergence. Thus
\begin{align*}
    \int \sqrt{g} \LL d^{4}x &
	= -\int \sqrt{g} Y^{\mu} [\nabla_{\mu},\nabla_{\nu}] Y^{\nu} 
	    + \ldots = \\
	& = \int \sqrt{g} Y^{\mu} S^{\nu}_{\beta,\mu\nu} Y^{\beta}
	    + \ldots = \\
	& = \int \sqrt{g} Y^{\mu} g^{\nu \alpha} S_{\alpha\beta,\nu\mu}
	    Y^{\beta} + \ldots ,
\end{align*}
    where we do not write explicitly contribution of torsion,
    and so to realize the main proposal we need condensate
\begin{equation*}
    <Y^{\mu} Y^{\beta}> \sim \frac{1}{\kappa^{2}} g^{\mu\beta} 
\end{equation*}
    which will enter also the kinetic energy of
$ Z_{\mu\alpha} $.
    So it seems, that it is preferable to use the lagrangian
(\ref{LXf})
    per se and interprete it as a kinetic term for vector fields
$ f_{\mu}^{A} $.

\end{document}